\documentclass[12pt]{article}
\usepackage{epsfig, amssymb}
\usepackage{graphicx,epsfig} 
\begin{document}
\begin{titlepage}
\thispagestyle{empty}

\bigskip

\begin{center}
\noindent{\Large \textbf
{Analytic Approaches to anisotropic Holographic Superfluids}}\\
\vspace{2cm} \noindent{Pallab Basu ${}^{a,b}$\footnote{e-mail: pallabbasu@gmail.com }and
Jae-Hyuk Oh${}^{a,b,c}$\footnote{e-mail: jack.jaehyuk.oh@gmail.com}}

\vspace{1cm}
  {\it
 Department of Physics and Astronomy, \\
 University of Kentucky, Lexington, KY 40506, USA ${}^a$\\
 
\vspace{0.2cm}
 }
 {\it Department of Physics and Center for Quantum Spacetime, \\
Sogang University, Seoul 121-742, South Korea ${}^b$\\
 }
\vspace{0.2cm}
  {\it
 Harish-Chandra Research Institute, \\
 Chhatnag Road, Jhunsi, Allahabad-211019, India ${}^c$\\
 }
\end{center}

\vspace{0.3cm}
\begin{abstract}
We construct an analytic solution of the Einstein-$SU(2)$-Yang-Mills system as the holographic dual of an anisotropic superfluid near its critical point, 
up to leading corrections in both the inverse Yang-Mills coupling and a symmetry breaking order parameter. We have also calculated the ratio of shear viscosity 
to entropy density in this background, and shown that the universality of this ratio  is lost in the broken symmetry direction. 
The ratio displays a scaling behavior near the critical point with critical exponent 
$\beta=1$, at the leading order in the double expansion.
\end{abstract}
\end{titlepage}

\newpage

\tableofcontents
\newpage

\section{Introduction}

Gauge/gravity duality has led to many useful insights into strongly coupled field theories. Recently, 
fluid/gravity duality has been widely studied, providing useful information about conformal fluid dynamics, 
the effectively long wavelength description of conformal field theory. Many crucial quantities characterizing conformal fluids can be  obtained as real-time Green's functions in dual gravitational theories \cite{Son1,Son2}.
 
The most celebrated example from holographic fluid dynamics is the ratio of shear viscosity $\eta$ to entropy density
$s$, $\frac{\eta}{s}=\frac{1}{4\pi}$ \cite{Son3,Alex1,Alex2,Liu1}. The ratio seems to be  universal for theories with weakly coupled gravity duals
\footnote{This universality can be violated when string effects or quantum effects are to be taken into account in dual gravity\cite{Alex3,Alex4,Aninda1}.}.
However, in recent studies of anisotropic conformal fluids \cite{Basu:2009vv,Johanna1},  the universality of the ratio turns out to be violated
\cite{Johanna2,Natsuume:2010ky}. These studies have considered an exact solution of the bulk Einstein-$SU(2)$-Yang-Mills action
\footnote{See \cite{Manvelyan:2008sv,Gubser:2008zu,Gubser:2008wv,Roberts:2008ns} for pioneering works on connection between Einstein-$SU(2)$-Yang-Mills and $p$-wave holographic superfluids.} for an $AdS$ black brane in 5 dimensions with nonzero chemical potential.  Corresponding to the chemical potential, the temporal component of the gauge potential is proportional to $\sigma_3$ of the $SU(2)$ gauge group. The boundary metric enjoys 
$SO(3)$ global symmetry (symmetry under spatial rotations). At high temperature (equivalently small chemical potential $\mu$), the system stays in the
isometric phase. However at a certain chemical potential $\mu=\mu_c$, 
the $SO(3)$ symmetry is {\it spontaneously} broken to $SO(2)$ because
one of the spatial components of the Yang-Mills field develops a non-trivial zero mode (as a
solution of the linearized Yang-Mills field equations) and it is normalizable solution. 
It turns out that in the region of $\mu>\mu_c$, this mode condenses and there is a new anisotropic superfluid phase.

As long as we are looking at a solution with  $SO(3)$ symmetry, the universality of the ratio of entropy density and shear viscosity holds. 
{\it This is because the shear viscosity depends on  gravitational perturbations in tensor modes of $SO(3)$ in the dual gravity. }
Each tensor mode satisfies a massless scalar field equation decoupled from
the others, which ensures universality. The anisotropic symmetry-broken phase provides an anisotropic shear viscosity to entropy density ratio. The reason is that once $SO(3)$ symmetry is broken to $SO(2)$, the gravitational wave modes in the 
broken-symmetry directions are no longer $SO(2)$ tensor modes. They are not decoupled
from other fields and in fact interact with Yang-Mills fields. Therefore, in this case the gravitational modes do not display universal behavior.

Near the critical point($\mu = \mu_c$), the phase transition is expected to depend on 
$\alpha^2 \equiv \frac{\kappa^2_{5}}{g^2}$\cite{Johanna2}, 
where $\kappa_5$ is 5-dimensional gravity constant and $g$ is Yang-Mills coupling. If $\alpha$ is less
than a certain critical value $\alpha_{crit}$, 
the phase transition becomes second order. Near the second order phase transition, the ratio displays a scaling behavior with some critical exponent $\beta$,
\begin{equation}
\label{scaling BEHAVIOR}
1-4\pi\frac{\eta}{s} \sim \left( 1-\frac{T}{T_c} \right)^{\beta},
\end{equation}
where the value $\beta = 1.00 \pm 0.03$ has been calculated numerically, taking 
into account the back reaction to the background metric.  
For $\alpha > \alpha_{crit}$, the phase transition is 
first order. 
Unfortunately, the large-$\alpha$ region has not been 
explored very well numerically, due to technical difficulties. 
Another interesting remark from \cite{Johanna2} is that the critical exponent, $\beta$ does not seem to depend on $\alpha$.

To examine such properties rigorously an analytic
approach is needed.  Our starting point for such an approach is a zero-mode solution at zero gravitational coupling, which is known exactly at the critical point \cite{Basu1}. In this note, we perturbatively analyze the properties of the anisotropic fluid near the critical point.  
We obtain the back reaction to the space time metric
and also solve linearized equations of the gravitational perturbations and Yang-Mills fields from the back-reacted metric, using a double
expansion of inverse Yang-Mills coupling $\alpha^2$ and $\varepsilon \tilde{D}_1$. 
$\varepsilon$ is a dimensionless small parameter and $\tilde{D}_1$ is an $SO(3)$ symmetry-breaking scale appearing in the anisotropic part of the Yang-Mills field. 
Similar perturbative expansions for different holographic models have been discussed by various authors \cite{Herzog1,Herzog2,Bhattacharya:2011tra,Bhattacharya:2011ee}. 
However unlike those works, we consider both an anisotropy and the gravitational back reaction from the Yang-Mills fields.

In our double expansion, the nontrivial leading order turns out to be $O(\alpha^2 \varepsilon^2)$. We also get the shear viscosity and entropy density ratio up to this order. 
Nonuniversality of the ratio of shear viscosity to entropy density shows up
 in the directions of broken symmetry while the ratio in the unbroken direction displays the expected universality. In our perturbative analysis we do not see the first order phase transition because  $\alpha$ 
is small in the perturbative regime. Near the critical point, our solution also presents the scaling behavior 
as Eq(\ref{scaling BEHAVIOR}) and it turns out that the critical exponent $\beta = 1$ up to the leading order correction, 
which is consistent with the numerical result in \cite{Johanna2}. In general, our perturbative results agree with and complement the numerical results in \cite{Johanna2}.

\section{Holographic Setup and Large Coupling Expansion}
We consider the Einstein-$SU(2)$ Yang-Mills system in asymptotically $AdS_{5}$ spacetime. The action is
\begin{equation}
S=\int d^{5}x \sqrt{-G} \left(  \frac{1}{\kappa^2_5}(R+\frac{12}{L^2}) - \frac{1}{4g^2}F^{a}_{MN}F^{aMN} \right),
\end{equation}
where $M$, $N$... are 5-dimensional space-time indices, $a$.. are $SU(2)$ indices and $g$ is the Yang-Mills coupling. We conventionally choose $L=1$. The Yang-Mills field strength $F^{a}_{MN}$ is given by
\begin{equation}
F^{a}_{MN}=\partial_{M}A^{a}_{N}-\partial_{N}A^{a}_{M}-\epsilon^{abc}A^{b}_{M}A^{c}_{N},
\end{equation}
where $\epsilon^{abc}$ is anti-symmetric tensor with $\epsilon^{123}=1$.
The equations of motion from the action are obtained as
\begin{eqnarray}
\label{W-equation}
W_{MN}&\equiv&R_{MN}+4G_{MN}-\kappa^{2}_{5}\left( T_{MN}-\frac{1}{3} T^{P}_{P}G_{MN} \right)=0, \\ 
\label{Y-equation}
Y^{aN}&\equiv&\nabla_{M}F^{aMN}-\epsilon^{abc}A^{b}_{M}F^{cMN}=0,
\end{eqnarray}
where $T_{MN}$ is energy-momentum tensor, of which form is
\begin{equation}
T_{MN}=\frac{1}{g^2}\left( F_{MP}^{a}F_{N}^{Pa} -\frac{1}{4}F_{PQa}F^{PQa}G_{MN} \right).
\end{equation}
Our ansatz for the metric and Yang-Mills field are given by
\begin{eqnarray}
\label{gauge and metric ansatz}
A&=&\phi(r)\tau^{3}dt+\omega(r)\tau^{1}dx, \\ \nonumber
ds^2&=&-N(r)\sigma^2(r)dt^2+\frac{dr^2}{N(r)}+r^2f^{-4}(r)dx^2+r^2f^{2}(r)\left( dy^2 + dz^2 \right),
\end{eqnarray}
where $\tau^{a}=\frac{s^{a}}{2}$ and $s^{a}$ are Pauli-matrices. 
The Yang-Mills field equations of motion in terms of the above ansatz are 
\begin{eqnarray}
\frac{r^2Y^{1}_{y}}{f^{4}(r)N(r)}&=&\omega^{\prime\prime}(r)+\left( \frac{1}{r}+\frac{\sigma^{\prime}(r)}{\sigma(r)}+\frac{N^{\prime}(r)}{N(r)} +4\frac{f^{\prime}(r)}{f(r)} \right) \omega^{\prime}(r) +\frac{\phi^{2}(r)\omega(r)}{N^{2}(r)\sigma^{2}(r)}=0, \\ \nonumber
\sigma^2(r)Y^{3}_{t}&=&\phi^{\prime\prime}(r)+\left( \frac{3}{r} - \frac{\sigma^{\prime}(r)}{\sigma(r)} \right)\phi^{\prime}(r) - \frac{f^{4}(r)\omega^{2}(r)}{r^{2}N(r)}\phi(r)=0,
\end{eqnarray}
and the Einstein equations are
\begin{eqnarray}
\frac{2W_{tt}}{N^2 (r)\sigma^2 (r)}&=&2\frac{\sigma^{\prime\prime}(r)}{\sigma(r)}+\frac{6}{r}\frac{\sigma^{\prime}(r)}{\sigma(r)}+\frac{N^{\prime\prime}(r)}{N(r)}+\frac{3}{r}\frac{N^{\prime}(r)}{N(r)}-\frac{8}{N(r)}+3\frac{\sigma^{\prime}(r)}{\sigma(r)}\frac{N^{\prime}(r)}{N(r)}\\ \nonumber
&-& \frac{2\kappa^{2}_{5}}{3g^2}\left( \frac{f^4(r)\omega^{\prime2}(r)}{r^2} +\frac{2\phi^{\prime 2}(r)}{N(r)\sigma^{2}(r)}+\frac{2f^{4}(r)\phi^{2}(r)\omega^2(r)}{r^2 \sigma^2(r)N^2(r)} \right)=0        \\
\bar{W}&\equiv&\frac{2r}{\sigma^{2}(r)N(r)}W_{tt}+2r^{2}N(r)W_{rr}\\ \nonumber
&=&-12r\frac{f^{\prime 2}(r)}{f^{2}(r)}+6\frac{\sigma^{\prime}(r)}{\sigma(r)}-\frac{2\kappa^{2}_{5}f^{4}(r)}{g^{2}r}\left( \omega^{\prime 2}(r) + \frac{\phi^{2}(r)\omega^{2}(r)}{N^{2}(r)\sigma^{2}(r)}  \right)=0, \\ 
\tilde{W}&\equiv&2W_{yy}+f^{6}(r)W_{xx} \\ \nonumber
&=&2-\frac{4r^2}{N(r)}+r\frac{N^{\prime}(r)}{N(r)}+r\frac{\sigma^{\prime}(r)}{\sigma(r)}+\frac{r^2 \kappa^2_{5}\phi^{\prime 2}(r)}{3g^2\sigma^{2}(r)N(r)}=0, \\ 
\frac{f^{4}(r) W_{xx}}{2r^2 N(r)}&=&\frac{f^{\prime\prime}(r)}{f(r)}+\left( \frac{\sigma^{\prime}(r)}{\sigma(r)}+\frac{3}{r} +\frac{N^{\prime}(r)}{N(r)} \right)\frac{f^{\prime}(r)}{f(r)}-\frac{1}{r^2}+\frac{2}{N(r)}-\frac{N^{\prime}(r)}{2rN(r)} -\frac{\sigma^{\prime}(r)}{2r\sigma(r)} \\ \nonumber
&-&\frac{\kappa^2_{5}}{g^2} \left( \frac{\omega^{\prime 2}(r)f^4(r)}{3r^2} - \frac{\phi^{2}(r)\omega^{2}(r)f^4(r)}{3r^2 \sigma^2 (r) N^2 (r)} + \frac{\phi^{\prime 2}(r)}{6N(r)\sigma^{2}(r)} \right)-\frac{f^{\prime 2}(r)}{f^2 (r)}=0 \\
W_{yy}&=&W_{zz}.
\end{eqnarray}
A known exact solution of the equations of motion is the AdS charged-black-brane solution given by \begin{eqnarray}
\label{the zeroth order in varepsilon}
\phi(r)&=&\tilde{\mu} (1-\frac{r^{2}_{h}}{r^2}), {\ \ }\omega(r)=0, \\ \nonumber
\sigma(r)&=&f(r)=1 {\rm \ \ and\ }N(r) = N_{0}(r) \equiv  r^2 -\frac{m}{r^2}+\frac{2\tilde{\mu}^2\alpha^2r^4_{h}}{3r^4} ,
\end{eqnarray}
where $\tilde{\mu}$ is chemical potential, $r_h$ is the black brane horizon and $m\equiv r^4_{h}+\frac{2\mu^2\alpha^2r^2_{h}}{3}$. In the infinite Yang-Mills coupling limit as $g \rightarrow \infty$, the last term in $N(r)$ vanishes and the solution becomes uncharged.

\subsection{Large Coupling Expansion and its Leading Order Corrections}
In this section, we develop 
corrections to the metric and Yang-Mills field perturbatively in a double expansion 
in $\varepsilon \tilde{D}_{1}$ and $\alpha^2 \equiv \frac{\kappa^{2}_{5}}{g^2}$. 
$\varepsilon$ is a dimensionless small parameter and $\tilde{D}_{1}$ is the $SO(3)$ rotational symmetry-breaking order parameter. 
We choose the horizon of the black brane to be conventionally located at $r=1$ 
by scaling  $r \rightarrow r_{h} r$ and $\{t,x,y,z \} \rightarrow \frac{1}{r_{h}}\{t,x,y,z \}$ 
and defining a new chemical potential $\mu \equiv \frac{\tilde{\mu}}{r_{h}}$.
The equations of motion enjoy a certain scaling symmetry\cite{Johanna1,Basu:2009vv}. 
By means of these rescalings, 
we can choose the asymptotic values of $\sigma(r=\infty)=1$ and $f(r=\infty)=1$ 
at the large $r$ boundary where the spacetime becomes asymptotically $AdS_{5}$. 
The value of the chemical potential in the dual boundary field theory  is taken to be $\mu=4$ 
at the phase transition point. To obtain corrections, we expand any fields $a(r)$ appearing 
in the ansatz (\ref{gauge and metric ansatz}) as
\begin{equation}
\label{varepsilon expansion}
a(r)=a_{0}(r)+\varepsilon a_{1}(r) +\varepsilon^2 a_{2}(r)...
\end{equation}
Each term in the expression can in turn be expanded as
\begin{equation}
\label{alpha expansion}
a_{i}(r)=a_{i,0}(r)+\alpha^2a_{i,2}+\alpha^4a_{i,4}(r)...
\end{equation}
The zeroth-order solution in $\varepsilon$ is given in Eq(\ref{the zeroth order in varepsilon}), 
where only $N_{0}$ contains a subleading correction of order
$\alpha^2$ in the sense of the above expansion. $N_{0,2}=\frac{32}{3}\left( \frac{1}{r^4}-\frac{1}{r^2} \right)$ and the higher-order terms in $\alpha^2$ vanish, $N_{0,i}=0$ for $i=4,6...$. 
The detailed computations of the nontrivial leading order corrections to metric and Yang-Mills field are given 
in Appendix \ref{Leading Order Solutions}. 
Here, we briefly list the leading-order back-reaction corrections to the metric, which are given by
\begin{eqnarray}
\label{background_metric_ea}
\sigma(r)&=&1-\varepsilon^2\alpha^2\frac{2\tilde{D}^2_{1}}{9(1+r^2)^3}, {\ \ }f(r)=1-\varepsilon^2\alpha^2\frac{\tilde{D}^2_{1}(1-2r^2)}{18(1+r^2)^4} \\ \nonumber
{\rm and \ \ }N(r)&=&r^2-\frac{1}{r^2} +\frac{32\alpha^2}{3}\left( \frac{1}{r^4}-\frac{1}{r^2} \right)
-\varepsilon^2\alpha^2\frac{4\tilde{D}^2_{1}}{9r^2}\left( \frac{1+2r^2}{r^2(1+r^2)^3}-\frac{3r^2}{2(1+r^2)^2} \right.\\ \nonumber
&+&\left.\frac{281}{560}\left(1-\frac{1}{r^2}\right)\right).
\end{eqnarray}
Any subleading corrections to the Yang-Mills field in $\alpha^2$ would not contribute to the leading back-reaction corrections to the metric 
(Our aim is to get metric corrections up to $O(\alpha^2 \varepsilon^2)$).
Therefore, we obtain the Yang-Mills field solutions up to $\phi_{i,0}$ and $\omega_{i,0}$ only. These are given by
\begin{eqnarray}
\omega(r)&=&\varepsilon\frac{\tilde{D}_{1}r^2}{(r^2 +1)^2} + O(\varepsilon^2), \\ 
\phi(r)&=&4(1-\frac{1}{r^2})+\frac{\varepsilon^2\tilde{D}^2_{1}}{4}\left(\frac{(1+2r^2)}{3r^2 (1+r^2)^3}-\frac{1}{8}+\frac{281}{1680}\left(1-\frac{1}{r^2}\right)\right)
+O(\varepsilon^3).
\end{eqnarray}
The black brane temperature is modified by the leading corrections to
\begin{equation}
T=\frac{1}{\pi}\left( 1-\frac{16}{3}\alpha^2+\frac{17}{1260}\tilde{D}^2_{1}\varepsilon^2 \alpha^2 \right),
\end{equation}
where $T_{c} \equiv \frac{1}{\pi}\left( 1-\frac{16}{3}\alpha^2 \right)$ is the critical temperature at the phase transition 
from the isotropic phase to the anisotropic phase.
The black brane entropy is 
\begin{equation}
\label{Anisotrophic entropy}
S=\frac{2\pi}{\kappa^2_5}V_{3},
\end{equation}
where $V_{3}$ is spatial coordinate volume of the boundary space-time, $V_{3} = \int dxdydz$, in this rescaled coordinate.

\section{Anisotropy of Shear Viscosities}
In this section, we calculate the ratio of shear viscosity to entropy density via the Kubo formula, by considering fluctuations $h_{MN}$ and $\delta A^{a}_{M}$ around the background 
metric and the background Yang-Mills field, respectively. We choose the gauge $h_{Mr}=\delta A^{a}_{r}=0$. In the anisotropic phase, the bulk gravity system enjoys residual $SO(2)$ and $Z_2$ symmetries.  The modes may be decomposed according to their $SO(2)$ representations as 
\begin{itemize}
\item Scalar modes in $SO(2)$ : $h_{yz}$; $h_{yy}-h_{zz}$,
\item Vector modes in $SO(2)$ : $h_{yt}$,$\delta A^3_y$;  $h_{xy}$,$\delta A^{1}_{y}$,$\delta A^{2}_{y}$; $h_{zt}$, $\delta A^3_z$;  $h_{xz}$,$\delta A^{1}_{z}$,$\delta A^{2}_{z}$,
\item Tensor Modes in $SO(2)$ : $h_{tt}$,$h_{yy}+h_{zz}$,$h_{xx}$,$h_{xt}$,$\delta A^{a}_{t},\delta A^{a}_{x}$,
\end{itemize} 
where each decoupled mode is categorized by semicolons. $h_{yz}$ is totally decoupled from any other modes and satisfies a massless scalar field equation showing universality. 
However $h_{xy}$ interacts with $\delta A^{1}_{y}$ and $\delta A^{2}_{y}$, leading to nonuniversal behavior. In the following, we will obtain solutions for $h_{yz}$ and $h_{xy}$ 
and show this explicitly. Other modes can be calculated by the similar methods.
\subsection{Universality of $\frac{\eta_{yz}}{s}$}
In this subsection, we calculate the ratio of shear viscosity to  entropy density for the shear mode $h_{yz}$, using the double expansion 
that we introduced in the previous section. We consider fluctuations of the Yang-Mills field and metric fields around the
background metric(\ref{background_metric_ea}) and obtain perturbative corrections up to $O(\varepsilon^2 \alpha^2)$. 
We only consider time dependent fluctuations with frequency $\nu$ and use small frequency expansion up to first 
subleading order in $\nu$. Even in the presence of a nonzero symmetry breaking parameter $\tilde{D}_{1}$, the rotational 
symmetry in $y-z$ plane is not broken and the ratio $\frac{\eta_{yz}}{s}$ is universal. In the following, as a warm up, we will explicitly 
calculate this ratio to be $\frac{1}{4\pi}$ up to leading-order corrections in $\alpha^2 \varepsilon^2$.
To show this, we begin with the linearized equation of motion of $h_{yz}(r) \equiv r^2 f^2(r)\Phi(r,t)$,
\begin{equation}
\label{Phi equation}
0=\Phi^{\prime\prime}_{\nu}(r)+\left( \frac{1}{r}+\frac{4r}{N(r)}-\frac{\alpha^2 r \phi^{\prime 2}(r)}{3\sigma^2 (r)N(r)} \right)
\Phi^{\prime}_{\nu}(r)+\frac{\nu^2 \Phi_{\nu}(r)}{N^{2}(r)\sigma^2 (r)}, 
\end{equation}
where the prime denotes the radial derivative. For the field $\Phi(r)$, we have used the Fourier transform from real time to frequency, as
\begin{equation}
\Phi(r,t)=\int^{\infty}_{-\infty}e^{-i\nu t}\Phi_{\nu}(r) d\nu
\end{equation}

The near horizon behavior of $\Phi_{\nu}(r)$ should be a purely ingoing solution 
\begin{equation}
\label{purely ingoing}
\Phi_{\nu}(r)\sim \left( 1-\frac{1}{r} \right)^{-i\frac{\nu}{4}\left( 1+\frac{16}{3}\alpha^2 -\frac{17}{1260}\varepsilon^2 \alpha^2 
\tilde{D}^{2}_{1}\right)+O(\varepsilon^k \alpha^l)},
\end{equation}
where $k$ and $l$ are integers with $k>2$ or $l>2$. 
With this boundary condition, the solution $\Phi_{\nu}(r)$ is obtained as
\begin{equation}
\label{the general Phi solution}
\Phi_{\nu}(r)=\left( \frac{N(r)}{r^2} \right)^{-i\frac{\nu}{4}\left( 1+\frac{16}{3}\alpha^2 -\frac{17}{1260}\varepsilon^2 \alpha^2 
\tilde{D}^{2}_{1}\right)+O(\varepsilon^k \alpha^l)} F(\varepsilon, \alpha^2),
\end{equation}
where 
\begin{equation}
\label{the general F solution}
F(\varepsilon, \alpha^2)=\sum_{i,j=0}^{\infty} \Phi_{i,2j}(r)\varepsilon^{i} \alpha^{2j}.
\end{equation}
Each $\Phi_{i,2j}$ and its near-$AdS$ boundary expansion are given in Appendix \ref{Leading Order Perturbation of h_yz}.
Here, we briefly discuss the near boundary expansion of the solution to get $\frac{\eta_{yz}}{s}$. 
Defining the boundary value $\tilde{\Phi} \equiv \Phi_{\nu}(\infty)$, $\Phi_{\nu}(r)$ can be expanded as
\begin{equation}
\Phi_{\nu}(r \rightarrow \infty)=\tilde{\Phi} + \frac{i\nu}{4 r^4} \tilde{\Phi} +O(r^{i}\nu^j \varepsilon^k \alpha^l)
\end{equation}
in the large $r$ limit (See Eq(\ref{near boundary expansion of hyz}) in Appendix \ref{Leading Order Perturbation of h_yz}), where
$i<-4$, $j>1$,$k>2$ or $l>2$.
Using the prescription to get retarded green's function in \cite{Johanna2}, we get
\begin{equation}
G^{R}_{yz,yz}(\nu,\vec{k}=0)=\frac{-i\nu}{2 \kappa^{2}_{5}}+O(\nu^2)
\end{equation}
The shear viscosity in $y-z$ direction is given by
\begin{equation}
\eta_{yz} \equiv \lim_{\nu \rightarrow 0} \frac{1}{2\nu i}[ G^{R\star}_{yz,yz}-G^{R}_{yz,yz}]=\frac{1}{2\kappa^2_{5}},
\end{equation}
where star indicates complex conjugate.
Using the entropy of the black brane(\ref{Anisotrophic entropy}), the ratio of shear viscosity to entropy density is obtained as
\begin{equation}
\frac{\eta_{yz}}{s}=\frac{1}{4\pi}.
\end{equation}
This value turns out to be universal up to $O(\varepsilon^2 \alpha^2)$.

\subsection{Nonuniversality of $\frac{\eta_{xy}}{s}$}
\label{Nonuniversality of etaxys}
We start with a set of equations with $h_{xy} \equiv r^2 f^2(r) \Psi(r,t)$, $\delta A^{1}_{y}$ and $\delta A^{2}_{y}$. 
The superscripts on $\delta A$ fields note $SU(2)$ indices and subscripts do space-time indices.
 We also solve these equations
in the frequency space by Fourier transform as in the previous subsection. 
The equations in the frequency space are given by
\begin{eqnarray}
\label{Psi-equation}
0&=&\Psi^{\prime\prime}(r)+\left( \frac{1}{r} +\frac{4r}{N(r)} +\frac{6f^{\prime}(r)}{f(r)} -\frac{r\alpha^2 \phi^{\prime 2}(r)}{3N(r)\sigma^{2}(r)} \right)\Psi^{\prime}(r) +\frac{\nu^2\Psi(r)}{N^{2}(r)\sigma^{2}(r)} \\ \nonumber
&+& \frac{2\alpha^2}{r^2 f^{2}(r)}\left( \omega^{\prime}(r)\delta A_{y}^{1\prime}(r) -\frac{\omega(r)\phi^{2}(r)\delta A_{y}^{1}(r) }{N^{2}(r)\sigma^{2}(r)} +\frac{i\nu\omega(r)\phi(r)\delta A_{y}^{2}(r) }{N^{2}(r)\sigma^{2}(r)}\right), \\
\label{a1-equation}
0&=&\delta A_{y}^{1\prime\prime}(r)+\left( \frac{1}{r} -\frac{2f^{\prime}(r)}{f(r)} +\frac{N^{\prime}(r)}{N(r)} 
+ \frac{\sigma^{\prime}(r)}{\sigma(r)} \right)\delta A_{y}^{1\prime}(r)
+ \left( \frac{\nu^2+\phi^{2}(r)}{N^{2}(r)\sigma^{2}(r)}  \right)\delta A_{y}^{1}(r)\\ \nonumber
&-&f^{6}(r)\omega^{\prime}(r)\Psi^{\prime}(r)-\frac{2i\nu \phi(r)\delta A_{y}^{2}(r)}{N^{2}(r)\sigma^{2}(r)}, \\
\label{a2-equation}
0&=&\delta A_{y}^{2\prime\prime}(r)+\left( \frac{1}{r} -\frac{2f^{\prime}(r)}{f(r)} 
+\frac{N^{\prime}(r)}{N(r)} + \frac{\sigma^{\prime}(r)}{\sigma(r)} \right)\delta A_{y}^{2\prime}(r) 
+\left( \frac{\nu^{2}+\phi^{2}(r)}{N^{2}(r)\sigma^{2}(r)} \right)\delta A_{y}^{2} \\ \nonumber
&-&\frac{f^4(r)\omega^{2}(r)}{r^2 N(r)}\delta A_{y}^{2}  +\frac{i\nu\phi(r)}{N^{2}(r)\sigma^{2}(r)}(-f^{6}(r)\omega(r)\Psi(r)+2\delta A_{y}^{1}(r)).
\end{eqnarray}
We expand each field with the same fashion as Eq(\ref{the general Phi solution}):
\begin{eqnarray}
\label{PsiA1A2-expansion}
\Psi(r)&=&\left( \frac{N(r)}{r^2} \right)^{-i\frac{\nu}{4}\left( 1+\frac{16}{3}\alpha^2 -\frac{17}{1260}\varepsilon^2 \alpha^2 
\tilde{D}^{2}_{1}\right)+O(\varepsilon^k \alpha^l)} G(\varepsilon, \alpha^2), \\ \nonumber
\delta A^{1}_{y}(r)&=&\left( \frac{N(r)}{r^2} \right)^{-i\frac{\nu}{4}\left( 1+\frac{16}{3}\alpha^2 -\frac{17}{1260}\varepsilon^2 \alpha^2 
\tilde{D}^{2}_{1}\right)+O(\varepsilon^k \alpha^l)} H(\varepsilon, \alpha^2), \\ \nonumber
\delta A^{2}_{y}(r)&=&\left( \frac{N(r)}{r^2} \right)^{-i\frac{\nu}{4}\left( 1+\frac{16}{3}\alpha^2 -\frac{17}{1260}\varepsilon^2 \alpha^2 
\tilde{D}^{2}_{1}\right)+O(\varepsilon^k \alpha^l)} I(\varepsilon, \alpha^2),
\end{eqnarray}
where the functions $G(\varepsilon, \alpha^2)$, $H(\varepsilon, \alpha^2)$ and $I(\varepsilon, \alpha^2)$ are expanded as
\begin{eqnarray}
\label{GHI-expansion}
G(\varepsilon, \alpha^2)&=&\sum_{i,j=0}^{\infty} \Psi_{i,2j}(r)\varepsilon^{i} \alpha^{2j}, \\ \nonumber
H(\varepsilon, \alpha^2)&=&\sum_{i,j=0}^{\infty} \delta A^{1}_{i,2j}(r)\varepsilon^{i} \alpha^{2j}, \\ \nonumber
{\rm and \ \ }I(\varepsilon, \alpha^2)&=&\sum_{i,j=0}^{\infty} \delta A^{2}_{i,2j}(r)\varepsilon^{i} \alpha^{2j}.
\end{eqnarray}

The solutions of the equations are listed in Appendix.\ref{Leading Order correction of hxy}. 
Corrections to the Yang-Mills fields which are subleadings in $\alpha^2$ do not contribute to the leading order corrections of the
shear viscosity, so we get $\delta A^{1}_{i,0}(r)$ and $\delta A^{2}_{i,0}(r)$ only(See Eq(\ref{Psi-equation})). 
We also use small frequency expansion as in the last subsection, and obtain the solutions 
up to $O(\nu)$. We specify purely ingoing boundary conditions for
the fields, the form of which are the same as Eq(\ref{purely ingoing}). Here, we discuss the near boundary expansion of $\Psi(r)$, obtaining the retarded correlator of $h_{xy}$ and 
the shear viscosity to entropy density ratio, $\frac{\eta_{xy}}{s}$. 

The near-$AdS$ boundary expansions of $\Psi(r)$, 
$\delta A_{y}^{1}$ and $\delta A_{y}^{2}$ are given by
\begin{eqnarray}
\delta A^{1}_{y}(r=\infty)&=&-\frac{i\varepsilon\nu}{192}
\left( 6\tilde{A}^{(0)}_{1,0}-22\bar{A}^{(0)}_{1,0}-\tilde{D}_{1}\psi^{(0)}_{0,0} \right)+O(\varepsilon\nu^2), \\ \nonumber
\delta A^{2}_{y}(r=\infty)&=&\frac{i\varepsilon\nu}{192}
\left( 22\tilde{A}^{(0)}_{1,0}-6\bar{A}^{(0)}_{1,0}-11\tilde{D}_{1}\psi^{(0)}_{0,0} \right)+O(\varepsilon\nu^2),
\end{eqnarray}
and
\begin{eqnarray}
\Psi(r)&=&(\psi^{(0)}_{0,0}+\varepsilon\psi^{(0)}_{1,0}+\varepsilon^2 \psi^{(0)}_{2,0})
+\alpha^2(\psi^{(0)}_{0,2}+\varepsilon\psi^{(0)}_{1,2}+\varepsilon^2 \psi^{(0)}_{2,2}) \\ \nonumber
&+&\nu(\psi^{(1)}_{0,0}+\varepsilon\psi^{(1)}_{1,0}+\varepsilon^2 \psi^{(1)}_{2,0})
+\nu\alpha^2(\psi^{(1)}_{0,2}+\varepsilon\psi^{(1)}_{1,2}+\varepsilon^2 \psi^{(1)}_{2,2}) \\ \nonumber
&+&\frac{\nu}{r^4}\left( \frac{i}{4}(\psi^{(0)}_{0,0}+\varepsilon\psi^{(0)}_{1,0}+\varepsilon^2 \psi^{(0)}_{2,0})
+ \frac{i\alpha^2}{4}(\psi^{(0)}_{0,2}+\varepsilon\psi^{(0)}_{1,2}+\varepsilon^2 \psi^{(0)}_{2,2}) \right. \\ \nonumber
&+& \left.\frac{i\alpha^2\varepsilon^2 \tilde{D}_{1}}{192}
(5\tilde{A}^{(0)}_{1,0}-11\bar{A}^{(0)}_{1,0} )+O(\nu^2 \varepsilon^3\alpha^3) \right).
\end{eqnarray}
The $SO(3)$ symmetry is broken {\it spontaneously}, so the Yang-Mills field should not provide any source terms to the dual 
field theory system. Therefore, $A^{1}_{y}(r)$ and $A^{2}_{y}(r)$ should become normalizable modes of the solutions, 
then we have 
\begin{equation}
 6\tilde{A}^{(0)}_{1,0}-22\bar{A}^{(0)}_{1,0}-\tilde{D}_{1}\psi^{(0)}_{0,0} =0,
 {\rm \ \ and \ \ } 22\tilde{A}^{(0)}_{1,0}-6\bar{A}^{(0)}_{1,0}-11\tilde{D}_{1}\psi^{(0)}_{0,0}=0.
\end{equation}
The solutions of these equations are
\begin{equation}
\label{normailzable condition}
\tilde{A}^{(0)}_{1,0}= \frac{59}{112}\tilde{D}_{1}\psi^{(0)}_{0,0}, {\rm \ \ and \ \ }
 \bar{A}^{(0)}_{1,0}= \frac{11}{112}\tilde{D}_{1}\psi^{(0)}_{0,0}.
\end{equation}
Using Eq.(\ref{normailzable condition}) and defining $\Psi(\infty) \equiv \Psi$ as a boundary value of $\Psi(r)$, 
the near boundary expansion of $\Psi(r)$ is given by
\begin{equation}
\Psi(r \rightarrow \infty)= \Psi + \frac{i\nu}{4r^4} \Psi +\varepsilon^2 \alpha^2 \nu \frac{29 i \tilde{D}^{2}_{1}\Psi}{3584r^4}. 
\end{equation}
The prescription of the retarded green's function in \cite{Johanna2} provides
\begin{equation}
G^{R}_{xy,xy}(\nu,\vec{k}=0)=\frac{-i\nu}{2 \kappa^{2}_{5}}\left( 1+ \frac{29}{896}\varepsilon^2 \alpha^2 \tilde{D}^2_{1}\right)+O(\nu^2)
\end{equation}
and the shear viscosity is calculated as
\begin{equation}
\eta_{xy} \equiv \lim_{\nu \rightarrow 0} \frac{1}{2\nu i}[ G^{R\star}_{xy,xy}-G^{R}_{xy,xy}]=\frac{1}{2\kappa^2_{5}}
\left( 1+ \frac{29}{896}\varepsilon^2 \alpha^2 \tilde{D}^2_{1}\right).
\end{equation}
Using entropy of the black brane(\ref{Anisotrophic entropy}), 
the ratio of shear viscosity and entropy density obtained as
\begin{equation}
\label{ratio eta_yz and s}
\frac{\eta_{xy}}{s}=\frac{1}{4\pi}\left( 1+ \frac{29}{896}\varepsilon^2 \alpha^2 \tilde{D}^2_{1}\right).
\end{equation}
Therefore, the shear viscosity and entropy ratio in $x-y$ direction is not universal, and we have shown this up to 
non-trivial leading order correction in $\alpha$ and $\varepsilon$.
Using the temperature of the black brane, Eq(\ref{ratio eta_yz and s}) can be written as
\begin{equation}
1-4\pi \frac{\eta_{xy}}{s}=\frac{1305\pi T_{c}}{544}\left( 1-\frac{T}{T_{c}} \right)^{\beta}, 
\end{equation}
where $\beta=1$. It is also shown that the critical exponent $\beta=1$ up to corrections of order $\varepsilon^2 \alpha^2$, near the phase transition point $T=T_c$.

\section*{Appendix}
\appendix
\section{Leading Order Solutions}
\label{Leading Order Solutions}
In this section, we solve Eq(\ref{W-equation}) and Eq(\ref{Y-equation}) to get leading order back reaction to the metric 
by symmetry breaking order parameter in Yang-Mills field, $\tilde{D}_{1}$. 
Using metric and Yang-Mills fields ansatz(\ref{gauge and metric ansatz}) and their expansions(\ref{varepsilon expansion}) 
and (\ref{alpha expansion}), we obtain the equations order by order in $\varepsilon$ and $\alpha$. 
Since the aim is to get leading back reaction to the metric, we  get Yang-Mills field solutions up to the zeroth order 
in $\alpha$ and the first order in $\varepsilon$ for $\omega(r)$ but the second order in $\varepsilon$ for $\phi(r)$, 
both of which provide first leading order corrections to the metric correctly. The first order Yang-Mills equations 
in $\varepsilon$ are given by
\begin{eqnarray}
0&=&\phi^{\prime\prime}_{1}(r)+\frac{3}{r}\phi^{\prime}_{1}(r)-\frac{8}{r^3}\sigma^{\prime}_{1}(r), \\ 
\label{omega1}
0&=&\omega^{\prime\prime}_{1}(r)+\left( \frac{1}{r}+ \frac{2(r^4+1)+r^3 N^{\prime}_{0}(r)}{r(r^4 -1)+r^3 N_{0}(r)} \right)\omega^{\prime}_{1}(r)+\frac{16\omega_{1}(r)}{(r^2 +1+\frac{r^2 N_{0}(r)}{r^2-1})^2},
\end{eqnarray}
and the first order Einstein equations in $\varepsilon$ are
\begin{eqnarray}
0&=&\sigma^{\prime}_{1}(r), \\ 
0&=&2N_{1}(r)+rN^{\prime}_{1}(r)+\frac{64\alpha^2}{3r^4}\left( \frac{r^3}{4}\phi^{\prime}_{1}(r)-2{\sigma_{1}(r)} \right), \\
0&=&2rN_{1}(r)+r^2N^{\prime}_{1}(r)+2(1-5r^4)f^{\prime}_{1}(r)+2r(1-r^4)f^{\prime\prime}_{1}(r) \\ \nonumber
&-&\frac{16\alpha^2}{3r^3(1+r^2)}\left( -2r^2(2N_{1}(r)+rN^{\prime}_{1}(r)) -8r(1+r^2-2r^4)f^{\prime}_{1}(r)-r^3(1+r^2)\phi^{\prime}_{1}(r)\right).
\end{eqnarray}

We expand every field appearing in above equations using Eq(\ref{alpha expansion}). With this expansion, the solutions of metric are trivial, which are given by
\begin{eqnarray}
\sigma_{1,0}(r)&=&\tilde{\sigma}_{1,0}, {\ \ }\sigma_{1,2}(r)=0, \\ \nonumber
f_{1,0}(r)&=&\tilde{f}_{1,0},{ \ \ }f_{1,2}(r)=0, \\ \nonumber
N_{1,0}(r)&=&\frac{\tilde{N}_{1,0}}{r^2}, {\ \ }N_{1,2}(r)=\frac{8}{3}\tilde{C_{1}} \left( \frac{1}{r^4}-\frac{\tilde{N}_{1,2}}{r^2} \right).
\end{eqnarray}
where we set $\tilde{\sigma}_{1,0}=\tilde{f}_{1,0}=0$ for the boundary values $\sigma(\infty)=f(\infty)=1$. $\tilde{N}_{1,0}=0$ and $\tilde{N}_{1,2}=1$ for the space-time has its horizon at $r=1$.
The Yang-Mills fields equations(\ref{omega1}) up to $O(\alpha^2)$ are given by 
\begin{eqnarray}
0&=&\omega^{\prime\prime}_{1,0}(r)+\frac{1+3r^4}{r(r^4 -1)}\omega^{\prime}_{1,0}(r)+\frac{16\omega_{1,0}(r)}{(r^2 +1)^2}, \\ \nonumber
0&=&\omega^{\prime\prime}_{1,2}(r)+\frac{1+3r^4}{r(r^4 -1)}\omega^{\prime}_{1,2}(r)+\frac{16\omega_{1,2}(r)}{(r^2 +1)^2}+\frac{128\tilde{D}_{1}(1+9r^2-2r^4)}{3r^2(1+r^2)^5}.
\end{eqnarray}
 The Yang-Mills field solutions are solved as
\begin{eqnarray}
\omega_{1,0}(r)&=&\frac{\tilde{D}_{1}r^2}{(r^2 + 1)^2}, \\ 
\phi_{1,0}(r)&=&\frac{\tilde{C}_{1}}{2}\left( 1-\frac{1}{r^2} \right), \\
\omega_{1,2}(r)&=&\frac{\tilde{D}_{2}r^2}{(r^2 + 1)^2}+\frac{128\tilde{D}_{1}r^2}{3(r^2 + 1)^2}\int^{r}_{\infty}dy \frac{(y^2+1)^3}{(y^2-1)y^3}\left(-\frac{136}{384}+\frac{1}{2}ln\left( \frac{2y^2}{y^2+1} \right)\right.\\ \nonumber
&+&\left.\frac{7+53y^2+43y^4+27y^6+6y^8}{12(1+y^2)^5} \right).
\end{eqnarray}
The second order equation of Yang-Mills field  in $\varepsilon$ is
\begin{equation}
0=\phi^{\prime\prime}_{2,0}(r)+\frac{3}{r}\phi^{\prime}_{2,0}(r)-\frac{8}{r^3}\sigma^{\prime}_{2,0}(r)-\frac{4r^2\tilde{D}^2_{1}}{(r^2 +1)^5},
\end{equation}
and Einstein equations are obtained as
\begin{eqnarray}
0&=&2N_{2}(r)+rN^{\prime}_{2}(r)+2\sigma^{\prime}_{2}(r)\left( r^2-\frac{1}{r^2}-\frac{32\alpha^2}{3r^2}+\frac{32\alpha^2}{3r^4} \right) \\ \nonumber
&+&\frac{64\alpha^2}{3r^4}\left( 
\frac{r^6}{64}\phi^{\prime 2}_{1}(r)+\frac{r^3}{4}\phi^{\prime}_{2}(r)
\right),\\
0&=&6\sigma^{\prime}_{2}(r)-\frac{8\alpha^2 \tilde{D}^{2}_{1}}{3}\frac{r}{(r^2 +1)^4}, \\
0&=&r(-1+r^4)f^{\prime\prime}_{2}(r)+(-1+5r^4)f^{\prime}_{2}(r)-\frac{4\alpha^2\tilde{D}^{2}_{1}}{3}\frac{r(r^2-1)(1-6r^2+r^4)}{(1+r^2)^5},
\end{eqnarray}
where the second order equation of $\omega(r)$ is not given, because which provides subleading corrections 
to the metric backreaction.

The Yang-Mills field solution of $\phi(r)$ in $O(\varepsilon^2)$ are given by
\begin{equation}
\phi_{2,0}(r)=\tilde{\phi}_{2,0}+\tilde{C}_{2}\left(1-\frac{1}{r^2}\right)
+\frac{(1+2r^2)\tilde{D}^{2}_{1}}{12r^2(1+r^2 )^{3}}
\end{equation}

and the metric corrections are
\begin{eqnarray}
\sigma_{2,0}&=&\tilde{\sigma}_{2,0}, \\ 
\sigma_{2,2}&=&\frac{2\tilde{D}^{2}_{1}}{9}\left( \tilde{\sigma}_{2,2} - \frac{1}{(1+r^2)^3} \right), \\ 
N_{2,0}(r)&=&\frac{\tilde{N}_{2,0}}{r^2}, \\
N_{2,2}(r)&=&-\frac{16}{3r^2}\left( \tilde{N}_{2,2}-\frac{\tilde{C}^2_{1}}{32r^2} - \frac{\tilde{C}_{2}}{r^2}+\frac{\tilde{D}^{2}_{1}}{12}\left(  \frac{1+2r^2}{r^2(1+r^2)^3} - \frac{3r^2}{2(1+r^2)^2}  \right)\right), \\
f_{2,0}(r)&=&\tilde{f}_{2,0}, \\
f_{2,2}(r)&=&\tilde{f}_{2,2}-\frac{(1-2r^2)\tilde{D}^2_{1}}{18(1+r^2)^4},
\end{eqnarray}
where $\tilde{\phi}_{2,0}=-\frac{\tilde{D}^{2}_{1}}{32}$ for the regularity of the Yang-Mills field at the horizon 
and $\tilde{N}_{2,0}$, $\tilde{f}_{2,0}$, $\tilde{\sigma}_{2,2}$ and $\tilde{f}_{2,2}$ are $O(1)$ constants 
which are set to be vanished for $f(\infty)=\sigma(\infty)=1$. 
$\tilde{N}_{2,2}=\frac{\tilde{C}^{2}_{1}}{32}+\tilde{C}_{2}$ for the space-time has its horizon at $r=1$. 
$\tilde{C}_{1}$ and $\tilde{C_{2}}$ are coefficients of zero modes of Yang-Mills field equations. 
Without loss of generality, we can set $\tilde{C}_{1}=0$. However, $\tilde{C_{2}}=\frac{281}{6720}\tilde D^2_1$ requesting $\omega_{3,0}(r)$ to be normalizable mode and
regular at the black brane horizon. 
Then, we have metric backreaction only with $\tilde{D}_{1}$ as a $SU(2)$ symmetry breaking scale. 

\section{Leading Order Correction of $h_{yz}$}
\label{Leading Order Perturbation of h_yz}
In this section, we briefly describe the solutions of Eq(\ref{Phi equation}) using the form of 
expansion(\ref{the general Phi solution})
 and (\ref{the general F solution}). Each term in expansion(\ref{the general F solution}) is given by
\begin{eqnarray}
\Phi_{0,0}(r)&=&\phi^{(0)}_{0,0}+\nu \phi^{(1)}_{0,0} + O(\nu^2), \\ 
\Phi_{0,2}(r)&=&\phi^{(0)}_{0,2} +\nu \left( \phi^{(1)}_{0,2} +8i\phi^{(0)}_{0,0}
\left( ln\left( 1+\frac{1}{r^2} \right) -\frac{1}{r^2}\right) \right)+O(\nu^2), \\
\Phi_{1,0}(r)&=&\phi^{(0)}_{1,0} +\nu \phi^{(1)}_{1,0}+ O(\nu^2),
 {\ \ } \Phi_{2,0}(r)=\phi^{(0)}_{2,0} +\nu \phi^{(1)}_{2,0}+ O(\nu^2), \\
\Phi_{1,2}(r)&=&\phi^{(0)}_{1,2} +\nu \left( \phi^{(1)}_{1,2}+8i\phi^{(0)}_{1,0}
\left( ln\left( 1+\frac{1}{r^2} \right) -\frac{1}{r^2}\right) \right)+O(\nu^2),
\end{eqnarray}
and
\begin{eqnarray}
\Phi_{2,2}(r)&=&\phi^{(0)}_{2,2}+\nu \phi^{(1)}_{2,2}-\frac{i\nu}{840}\left( \phi^{(0)}_{0,0}\tilde{D}^{2}_{1}\left( 
-\frac{105}{2(1+r^2)} + \frac{175}{2(1+r^2)^2} -\frac{70}{3(1+r^2)^3} \right. \right.  \\ \nonumber
&-&\frac{35}{(1+r^2)^4}-\left.\left.\frac{279}{2r^2} \right)+\frac{6720 \phi^{(0)}_{2,0}}{r^2} 
+192\left( \tilde{D}^{2}_{1}\phi^{(0)}_{0,0} -35 \phi^{(0)}_{2,0}\right)ln(1+\frac{1}{r^2})\right).
\end{eqnarray}
We also obtain the near $AdS$ boundary expansion of $\Phi(r)$ as
\begin{eqnarray}
\label{near boundary expansion of hyz}
\Phi_{\nu}(r \rightarrow \infty)&=&\Phi^{(0)}_{0,0}+\varepsilon \Phi^{(0)}_{1,0}+\varepsilon^2 \Phi^{(0)}_{2,0} 
+ \alpha^2 \left( \Phi^{(0)}_{0,2}+\varepsilon \Phi^{(0)}_{1,2}+\varepsilon^2 \Phi^{(0)}_{2,2} \right) \\ \nonumber
&+&\nu\left(\Phi^{(1)}_{0,0}+\varepsilon \Phi^{(1)}_{1,0}+\varepsilon^2 \Phi^{(1)}_{2,0} 
+ \alpha^2 \left( \Phi^{(1)}_{0,2}+\varepsilon \Phi^{(1)}_{1,2}+\varepsilon^2 \Phi^{(1)}_{2,2} \right)\right) \\ \nonumber
&+& \frac{i \nu}{4r^4}\left(\Phi^{(0)}_{0,0}+\varepsilon \Phi^{(0)}_{1,0}+\varepsilon^2 \Phi^{(0)}_{2,0} 
+\alpha^2 \left( \Phi^{(0)}_{0,2}+\varepsilon \Phi^{(0)}_{1,2}+\varepsilon^2 \Phi^{(0)}_{2,2} \right) \right) \\ \nonumber
&+&O(r^i \nu^j \varepsilon^k \alpha^l),
\end{eqnarray}
where $i<-4$, $j>1$, $k>2$ or $l>2$.

\section{Leading Order correction of $h_{xy}$}
\label{Leading Order correction of hxy}
In this section, we list the solutions of the set of equations(\ref{Psi-equation}), (\ref{a1-equation}) and (\ref{a2-equation}).
We listed our solution using the expansion(\ref{PsiA1A2-expansion}) and (\ref{GHI-expansion}).
As explained in Sec.\ref{Nonuniversality of etaxys}, we get $\delta A^{1}_{i,0}$ and $\delta A^{2}_{i,0}$ only 
for the Yang-Mills field solution. $\delta A^{1}_{0,0}$ and $\delta A^{2}_{0,0}$ are zero modes of the solutions. Without
loss of any generality, we set $\delta A^{1}_{0,0}=\delta A^{2}_{0,0}=0$. The first subleading corrections of Yang-Mills
fields in $\varepsilon$
are given by

\begin{eqnarray}\delta A^{1}_{1,0}
(r)&=&\frac{r^2}{(1+r^2)^2}\left( \tilde{A}^{(0)}_{1,0} +\nu\tilde{A}^{(1)}_{1,0} -\frac{i\nu}{192r^2}
\left( 6\tilde{A}^{(0)}_{1,0}
(1+r^4+24r^2ln(r)\right. \right.  \\ \nonumber 
&-&8r^2ln(1+r^2)) + 2\bar{A}^{(0)}_{1,0}(5-11r^4-40r^2ln(r)-24r^2ln(1+r^2))\\ \nonumber
&-&\left.\left.\tilde{D}_{1}\psi^{(0)}_{0,0}(17+r^4+56r^2ln(r)-24r^2ln(1+r^2))\right)+ O(\nu^2) \right),
\end{eqnarray}
and
\begin{eqnarray}
\delta A^{1}_{1,0}(r)&=&\frac{r^2}{(1+r^2)^2}\left( \bar{A}^{(0)}_{1,0} +\nu\bar{A}^{(1)}_{1,0} +\frac{i\nu}{192r^2}
\left( 2\tilde{A}^{(0)}_{1,0}
(-5+11r^4+40r^2ln(r) \right. \right.  \\ \nonumber
&+&24r^2ln(1+r^2)) - 6\bar{A}^{(0)}_{1,0}(1+r^4+24r^2ln(r)-8r^2ln(1+r^2))\\ \nonumber
&+&\left.\left.\tilde{D}_{1}\psi^{(0)}_{0,0}(5-11r^4-40r^2ln(r)-24r^2ln(1+r^2))\right)+ O(\nu^2) \right).
\end{eqnarray}
$\Psi(r)$ solution is also obtained as
\begin{eqnarray}
\Psi_{0,0}(r)&=& \psi^{(0)}_{0,0}+\nu \psi^{(1)}_{0,0}+O(\nu^2), \\ 
\Psi_{1,0}(r)&=&\psi^{(0)}_{1,0}+\nu \psi^{(1)}_{1,0}+O(\nu^2), \\ 
\Psi_{2,0}(r)&=& \psi^{(0)}_{2,0}+\nu \psi^{(1)}_{2,0}+O(\nu^2), \\
\Psi_{0,2}(r)&=& \psi^{(0)}_{0,2}+\nu\psi^{(1)}_{0,2}+ 8i\nu\psi^{(0)}_{0,0}\left( ln\left( 1+\frac{1}{r^2} \right)-\frac{1}{r^2} \right)
+O(\nu^2), \\ 
\Psi_{1,2}(r)&=&\psi^{(0)}_{1,2}+\nu \psi^{(1)}_{1,2}+ 8i\nu\psi^{(0)}_{1,0}\left( ln\left( 1+\frac{1}{r^2} \right)-\frac{1}{r^2} \right)
+O(\nu^2), 
\end{eqnarray}
and
\begin{eqnarray}
\Psi_{2,2}(r)&=&\psi^{(0)}_{2,2}+ \frac{\tilde{A}^{(0)}_{1,0} \tilde{D}_{1}}{(1+r^2)^4}
-\frac{2\tilde{A}^{(0)}_{1,0} \tilde{D}_{1}}{3(1+r^2)^3} \\ \nonumber
&+&\nu \psi^{(1)}_{2,2} 
+\frac{\nu}{576}\left( \frac{8i\tilde{D}_{1} ( -3 \tilde{A}^{(0)}_{1,0}+48i\tilde{A}^{(1)}_{1,0}
+17\bar{A}^{(0)}_{1,0}+2\tilde{D}_{1}\psi^{(0)}_{0,0})}{(1+r^2)^3} \right.\\ \nonumber
&+& \frac{2i\tilde{D}_{1}(3\tilde{A}^{(0)}_{1,0} -5\bar{A}^{(0)}_{1,0} +4\tilde{D}_{1}\psi^{(0)}_{0,0} )}{1+r^2}
- \frac{2i\tilde{D}_{1}(-9\tilde{A}^{(0)}_{1,0} +19\bar{A}^{(0)}_{1,0} +37\tilde{D}_{1}\psi^{(0)}_{0,0} )}{(1+r^2)^2} \\ \nonumber
&+& \frac{12\tilde{D}_{1}(3i\tilde{A}^{(0)}_{1,0}+48\tilde{A}^{(1)}_{1,0}-3i\bar{A}^{(0)}_{1,0}-i\tilde{D}_{1}\psi^{(0)}_{0,0})}{(1+r^2)^4}
+\frac{36i(93\tilde{D}^{2}_{1}\psi^{(0)}_{0,0}-4480\psi^{(0)}_{2,0})}{35r^2} \\
&-&\frac{8i\tilde{D}_1(2r^2-1)ln(r)}{(1+r^2)^4}( -18\tilde{A}^{(0)}_{1,0}+10\bar{A}^{(0)}_{1,0}+7\tilde{D}_{1}\psi^{(0)}_{0,0}) \\ \nonumber
&-&4i(-3\tilde{A}^{(0)}_{1,0}\tilde D_1+5\bar{A}^{(0)}_{1,0}\tilde D_1-\frac{1814}{35}\tilde{D}^2_{1}\psi^{(0)}_{0,0}+2304\psi^{(0)}_{2,0})ln(r) \\ \nonumber
&+&\frac{24i\tilde{D}_1(2r^2-1)ln(1+r^2)}{(1+r^2)^4}( -2\tilde{A}^{(0)}_{1,0}-2\bar{A}^{(0)}_{1,0}
+\tilde{D}_{1}\psi^{(0)}_{0,0}) \\ \nonumber
&+&iln(1+r^2)( -6\tilde{A}^{(0)}_{1,0}+10\bar{A}^{(0)}_{1,0} -\frac{3628}{35}\tilde{D}^2_{1}\psi^{(0)}_{0,0}+4608\psi^{(0)}_{2,0}) \\ \nonumber
&+&O(\nu^2).
\end{eqnarray}
\section*{Acknowledgements}
First of all, the authors thank Shesansu Sekhar Pal. He answered every stupid question from us politely. We would like to thank Yun-Seok Seo, Wonwoo Lee, Changhun Oh, Kwanghyun Jo and DongHan Yum for discussions. We would also like to thank to everyone at
CQUeST (Sogang University) for hospitality, especially Chan-yong Park , Bum-Hun Lee and Sang-Jin Sin. Finally, J.H.O  thanks to his $\mathcal W.J.$ 

J.H.O. was supported by a National Science Foundation grant NSF-PHY-0855614, 
the National Research Foundation of Korea (NRF) grant funded by the Korean government (MEST) through the Center for Quantum Spacetime (CQUeST) 
of Sogang University with grant number 2005-004940. The work of P.B. is partially supported by National Science Foundation grants PHY-0970069 and PHY-0855614.

\end{document}